\documentclass[12pt]{article}
\usepackage{setspace}
\usepackage[english]{babel}
\usepackage{pdfpages}
\usepackage{multicol}
\usepackage[letterpaper,top=2cm,bottom=2cm,left=1in,right=1in,marginparwidth=1in]{geometry}

\usepackage{amsmath}
\usepackage{graphicx}
\usepackage{subcaption}
\usepackage[colorlinks=true, allcolors=blue]{hyperref}

\usepackage[super,sort&compress,comma]{natbib}%[square]
\usepackage{makecell}
\usepackage{lscape}
\usepackage{caption}
\usepackage{multirow}
\usepackage{bm}

\definecolor{blue}{rgb}{0,0,1}

\newcommand\locrone[1]{}%{\printTodo{#1}}%
\newcommand{\revone}{}%\color{blue}}%

\usepackage{amsmath}
\DeclareMathOperator*{\argmax}{arg\,max}

\makeatletter

\newcommand*{\addFileDependency}[1]{% argument=file name and extension
\typeout{(#1)}% latexmk will find this if $recorder=0
\@addtofilelist{#1}
\IfFileExists{#1}{}{\typeout{No file #1.}}
}\makeatother

\title{A Bayesian Hybrid Design with Borrowing from Historical Study}
\author{Zhaohua Lu\thanks{Daiichi Sankyo Inc.},~%\thanks{zhlu@dsi.com}~,
        John Toso\footnotemark[1],~%\thanks{jtoso@dsi.com}~,
        Girma Ayele\footnotemark[1],~%\thanks{gayele@dsi.com}~,
        Philip He\footnotemark[1]~\thanks{philip.he@daiichisankyo.com}       
}
\date{}

\begin{document}
	\maketitle
\vspace{-1cm}
	\begin{abstract}
		In early phase drug development of combination therapy, the primary objective is to preliminarily assess whether there is additive activity from a novel agent when combined with an established monotherapy. Due to potential feasibility issues for conducting a large randomized study, uncontrolled single-arm trials have been the mainstream approach in cancer clinical trials. However, such trials often present significant challenges in deciding whether to proceed to the next phase of development {\revone due to the lack of randomization in traditional two-arm trials}. A hybrid design, leveraging data from a completed historical clinical study of the monotherapy, offers a valuable option to enhance study efficiency and improve informed decision-making. Compared to traditional single-arm designs, the hybrid design may significantly enhance power by borrowing external information, enabling a more robust assessment of activity. The primary challenge of hybrid design lies in handling information borrowing. We introduce a Bayesian dynamic power prior (DPP) framework with three components of controlling amount of dynamic borrowing. The framework offers flexible study design options with explicit interpretation of borrowing, allowing customization according to specific needs. Furthermore, the posterior distribution in the proposed framework has a closed form, offering significant advantages in computational efficiency. The proposed framework's utility is demonstrated through simulations and a case study.  \end{abstract}

	\newpage
 
	\section{Introduction}
 In early phase drug development of combination regimen, the primary objective is to preliminarily assess whether there is additive activity when a novel agent combined with an established monotherapy. Randomization \cite{1998_ICH,2000_Meldrum_Brief,2019_Grayling_Review} can effectively control confounding effects for known and unknown factors in assessing a new treatment's effectiveness, so it is recommended when practically feasible and clinically acceptable. However, in order to have adequate assessment, the traditional randomized controlled trial framework often requires a large size trial which may not be acceptable at early stage of cancer drug development. As a result, uncontrolled single-arm trials have become the common practice in pharmaceutical industry. However, significant challenges often arise due to lack of confounding control when evaluating diverse patient population in a small sized study. Externally controlled trials (ECTs), utilizing propensity score matching, can enhance data interpretation.
External control data as supporting information have been used in multiple regulatory submissions in European Medicines Agency (EMA) \citep{2023_Wang_Current} and Food and Drug Administration (FDA) \citep{2021_Jahanshahi_Use} in oncology and rare disease. FDA recently approved eflornithine in high-risk neuroblastoma based on an ECT study in December 2023 \citep{2023_FDA_eflornithine}. Thanks to the scientific breakthroughs, the standard of care in cancer treatments and medical practice change rapidly, which may lead to evolving clinical outcomes in current study compared to historical study \citep{2024_Struebing}. This can complicate the interpretation of the results based on ECTs \citep{2021_Jessica_Emulating}, which are associated with multiple inherit challenges in comparative analysis \citep{2023_FDA_ECA,2023_Wang_Current}. 
When considering whether to use an externally controlled trial design, we should determine whether it is possible to generate evidence capable of distinguishing the effect of the investigational treatment from outcomes attributable to the disease’s natural history, prognostic differences in the study populations, knowledge of treatment assignment, or other factors such as differences in concomitant therapies \citep{2023_FDA_ECA}.
 
Compared to ECTs, which solely utilize external data as controls, hybrid controlled trials (HCTs) \citep{2020_Zhu_hybrid,2020_Xu_Study,2022_Ventz_Design} integrate both external control data and concurrent control data within randomized controlled trials, which combine to form the hybrid control. This is especially relevant for combination therapy development when subject-level data are available from completed large historical monotherapy studies. 
 By incorporating external data, both HCTs and ECTs have potential advantages of shortened study duration, reduced sample size and minimized patient exposure to suboptimal treatments. However, HCTs, considering the similarity of external control data to concurrent control data, are generally more robust when incorporating external data \citep{2022_Ventz_Design}.
Moreover, HCT has an advantage of assessing the comparability and exchangeablity \citep{1976_Pocock_Combination} for the external control compared to concurrent control. 
HCTs have been considered in rare disease \citep{2023_vanEijk_Hybrid},  pediatric extrapolation \citep{2023_Spanakis}, oncology \citep{2022_Ventz_Design,2022_Li_Hybrid}, platform trials \citep{2022_Sonja_online,2021_Ren_statistical} with non-concurrent controls, vaccine development \citep{2020_Zhu_hybrid}, dose-finding \citep{2011_Yuan_bayesian,2021_Duan_Hi}, adaptive design \citep{2022_Lin_dynamic} and other advanced randomized trial designs \citep{2020_Lee_novel,2020_Neven_sample}.

Combining external control and concurrent control requires appropriate statistical handling of the potential variation between external control and concurrent control. The advantages gained from incorporating historical data can wane when inconsistencies arise without proper adjustment, potentially leading to increased bias, inflated type I error rates, and reduced statistical power \citep{2020_Xu_Study}. The impact of external control needs to be properly controlled, e.g., by down-weighting or discounting according to the similarity between concurrent control and external control \citep{2018_Lim_Minimizing}. However, it is challenging at the design stage to determine the amount of information borrowed from external control without knowing concurrent control and thus the similarity. Dynamic borrowing methods that account for the similarity have been shown to achieve satisfactory performance under certain scenarios and assumptions \citep{2018_Nikolakopoulos_Dynamic,2022_Lin_dynamic}.

Multiple methods for down weighting variation from historical data have been developed since the pioneering work of Pocock \citep{1976_Pocock_Combination}. Historical data are naturally considered as prior knowledge, and Bayesian methods are thus widely employed to construct informative prior with discounting approaches such as {\revone power priors and its variants \citep{2000_Ibrahim_Power,2015_Power_Ibrahim,2018_Nikolakopoulos_Dynamic,2006_Duan_normalized,2021_Carvalho_On},} 
%}, normalized power priors \citep{
commensurate priors \citep{2012_Brian_Commensurate,2018_Hong_Power}, meta-analytic predictive priors (MAP)\citep{2010_Neuenschwander_Summarizing,2014_rMAP_Schmidli}, network meta-analysis \citep{2011_Jones_Statistical} and Bayesian hierarchical priors \citep{2003_Thall_Hierarchical,2018_Chu_Bayesian} including mixture priors \citep{2016_Neuenschwander_Robust,2023_SAM_Yang}. Several frequentist methods have also been proposed in the literature. Participants are matched based on propensity score from experimental treatment arm, concurrent control arm and historical control arm to estimate average treatment effect \citep{2008_Elizabeth_Matching, 2019_Yuan_Design}. A conditional borrowing method is proposed to determine the borrowing weights based on similarity of the empirical distribution of log hazard ratio estimate obtained from re-sampling \citep{2022_Li_conditional}.
{\revone A dynamic power prior for binary endpoint in noninferiority trials\citep{2018_A_Liu} is proposed to down weight the historical control data according the the response rate of the study control.}
    \begin{figure}
    \centering
	\includegraphics[height=3.5in]{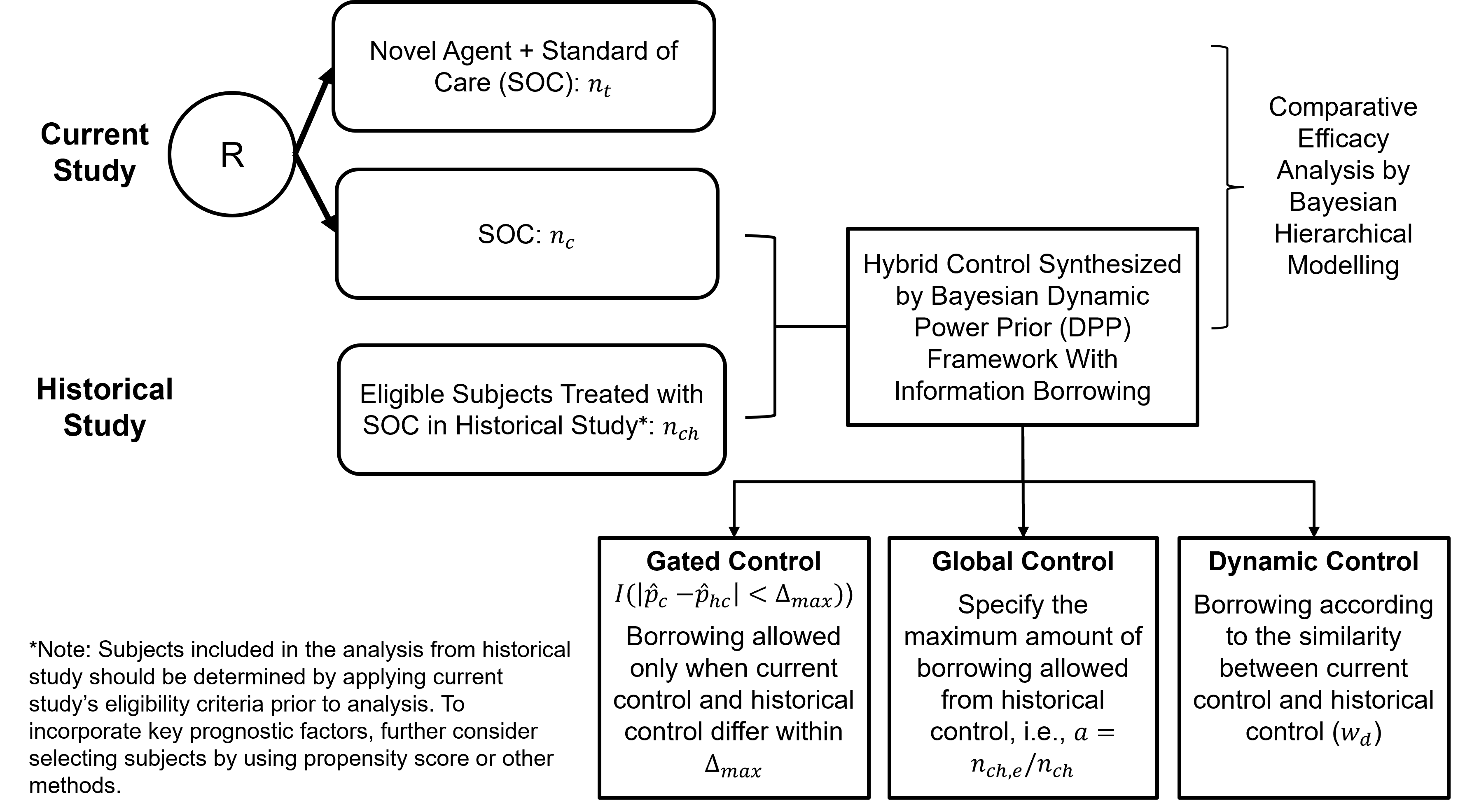}
	\caption{Bayesian Hybrid Design Using Dynamic Power Prior (DPP) Framework. {\revone The proposed DPP framework consists of three components: the gated control parameter, the global control parameter, and the dynamic control parameter. Together, these components effectively determine the amount of information borrowed from historical data to address prior-data conflict.}}
	\label{fig.BayesianHybridDesign}
    \end{figure}

%    \begin{figure}
%    \centering
%	\includegraphics[height=2.5in]{Decomposition.png}
%	\caption{Decomposition of Overall Borrowing in Bayesian Hybrid Design}
%	\label{fig.Decomposition}
%    \end{figure}

Despite of these methodology advancements, further developments are needed to better address certain practical challenges in implementation. Some existing methods for dynamic borrowing may be complex and lack of clear clinical interpretation of the components in statistical models and their utility in addressing  scenarios with different levels of prior-data conflict. Moreover, quantifying the amount of information borrowed from external control is not explicit. For the ease of understanding and endorsement of hybrid design by all stakeholders, the analytic formulation with clear interpretation will be particularly advantageous. In addition, computational intensive methods using Markov Chain Monte Carlo \citep{1984_Geman} can be time consuming and pose considerable hurdles when  exploring numerous design settings.

In this paper, we propose a Bayesian hybrid design framework based on dynamic power prior (DPP) to address these challenges.
The paper is organized as follows. In Section \ref{sec.BayesianDesign}, the proposed DPP framework is described that includes different components of borrowing control and it incorporates various methods of measuring similarity between concurrent control and historical control. Simulation studies in Section \ref{sec.results} demonstrated the value of each component in DPP borrowing with comparison to other methods. We applied the proposed hybrid design framework to a case study design and demonstrated its performance in Section \ref{sec.results}. Further discussions and practical considerations using hybrid design are provided in Section \ref{sec.discussion}.
 	
\section{Methods: A Bayesian Hybrid Design Framework}\label{sec.BayesianDesign}
Consider a hybrid study design as illustrated in Figure \ref{fig.BayesianHybridDesign}, which includes a randomized phase 2 study, referred to as the current study, along with some subjects from a previously completed historical study in the same patient population. Subjects in the current study are randomized to either experimental treatment or standard of care (SOC) treatment. A hybrid control is synthesized from the subjects randomized to the SOC treatment and subjects from the historical study who received the SOC treatment and met the eligibility criteria of the current study, referred to as the historical control. Consider tumor response (1 = Yes, 0 = No) as the primary endpoint in the hybrid study and denote $Y_{c}$, $Y_{t}$ and $Y_{ch}$ as the numbers of responses among $n_c$, $n_t$, and $n_{ch}$ subjects in the current study control, current study experimental arm and historical control respectively. Assume $Y_{t} \sim Binomial(n_t, p_t)$ and $Y_{c} \sim Binomial(n_c, p_c)$. An informative prior can be constructed from the historical control $p(p_c|Y_{ch}, {\bm \theta}_{c})\propto p(p_c|{\bm \theta}_{c})p(Y_{ch}|p_c)$, where $p_c$ is the response rate for control arm and ${\bm \theta}_{c}$ is the parameters for hyperprior $p(p_c|{\bm \theta}_{c})$. Then a hybrid control is synthesized by combining the informative prior and the concurrent control through the posterior distribution of $p(p_c|Y_{ch}, Y_{c}, {\bm \theta}_{c})$. The power prior method is utilized to control the amount of information borrowing through a weight parameter $w$, also called power parameter, so the informative prior of $p_c$ is adjusted as
$$p(p_c|Y_{ch}, {\bm \theta}_{c}) \propto p(p_c|{\bm \theta}_{c})\left\{p(Y_{ch}|p_c)\right\}^w,$$
where the power parameter $w\in [0, 1]$ controlling the amount of information borrowing from historical control. When the hyperprior $p(p_c|\mathbf{\theta_c})$ is a conjugate following a beta distribution $Beta(a_{0c}, b_{0c})$, then the posterior of $p_c$ follows a beta distribution 
  \begin{equation}\label{eq.posterior}
  p_c|Y_{ch}, Y_{c}, a_{0c}, b_{0c} \sim Beta(a_{0c} + Y_c + wY_{ch}, b_{0c} + (n_c-Y_c) + w(n_{ch}-Y_{ch})).    
  \end{equation}
  
Similarly, when the hyperprior $p(p_t|{\bm \theta_t})$ is a conjugate prior following a beta distribution $Beta(a_{0t}, b_{0t})$, the posterior of $p_t$ is $Beta(a_{ot} + Y_t, b_{0t} + (n_t-Y_t))$. One practical challenge is how to dynamically determine $w$ with customized control of borrowing as appropriate. We propose a dynamic power prior framework using Bayesian hierarchical model by decomposing the overall borrowing weight $w$ into 3 components:
    \begin{equation}
 	w = aw_dI(|\hat{p}_c-\hat{p}_{ch}|< \Delta_{max}), \label{eq:w}
    \end{equation}
  where $\hat{p}_{c}$ and $\hat{p}_{ch}$ are the observed response rate for control treatment group in the current study and the historical study respectively. The three components include: 
  \begin{enumerate}      
      \item Global borrowing parameter $a$. It is the maximum allowed amount of information that can be borrowed from historical control. For example, if we wish to borrow information equivalent to 40 subjects from a historical control of 200 subjects, then we set $a = \frac{n_{ch,e}}{n_c} = \frac{40}{200}$, where $n_{ch,e}$ represents the amount of information borrowing as equivalent to $n_{ch,e}$ subjects from the historical study. Please note that this is different from directly selecting $n_{ch,e}$ subjects from the historical study. This parameter controls the maximum amount allowed to borrow.   
      \item Gated borrowing parameter $\Delta_{max}$. It is the maximum tolerable threshold to determine whether information borrowing from historical control is allowed by comparing the empirical response rates observed in current control and historical control. No borrowing is allowed if the observed difference is greater than the pre-specified threshold, through the indicator function $I(|\hat{p}_c-\hat{p}_{ch}|< \Delta_{max})$. 
      \item Dynamic borrowing parameter $w_d$. Dynamic borrowing adaptively adjust the amount of borrowing according to the similarity between concurrent control and historical control. Multiple approaches can be considered and they are described in detail in Section \ref{sec.dynamicBorrowing}. 
    \end{enumerate}

  Each component is designed to address an important perspective in handling prior-data conflict. One can tailor the borrowing mechanism by optimizing the three borrowing components according to specific needs. The proposed dynamic borrowing framework provides an advantage of explicit data interpretation and customized control of borrowing. Furthermore, compared to {\revone normalized power priors} \citep{2006_Duan_normalized}, commensurate priors \citep{2012_Brian_Commensurate} and MAP \citep{2010_Neuenschwander_Summarizing} methods, it also has closed analytic formulas and does not require MCMC sampling, hence it is very efficient in computation for practical use. 
  %%%%%%%%%%%%%%%%%%%%%%%%
  
\subsection{Global Borrowing}
The proposed DPP provides a framework to incorporate various options to dynamically control the amount of information borrowing from the historical data. The global borrowing parameter $a$ controls the maximum amount of information that is allowed to borrow from historical control. Without considering the dynamic borrowing parameter $w_d$ and the gated borrowing parameter $\Delta_{max}$, the posterior distribution of $p_c$ is $ 	p(p_c|Y_{ch}, Y_{c}, a_{0c}, b_{0c})  \propto p(p_c|Y_{ch}, a_{0c}, b_{0c})p(Y_{c}|n_c, p_c)$ and hence it follows a beta distribution with parameters $Beta(a_{0c} + Y_c + aY_{ch}, b_{0c} + (n_c-Y_c)+a(n_{ch}-Y_{ch}))$. Then the posterior mean is calculated as 
\begin{equation} \label{eq.post.mean}
  \mu = \frac{{a_{0c} + Y_c + aY_{ch}}}{{a_{0c} +  b_{0c} + n_c + an_{ch}}}.    
\end{equation}
When $a$ is small ($\approx 0$) or the observed response rates of the concurrent control and historical control are similar $Y_c/n_c \approx Y_{ch}/n_{ch}$, the posterior mean $\mu$ is close to the posterior mean without borrowing from historical data
\begin{equation} \label{eq.post.mean.nohybrid}
\tilde{\mu} = ({a_{0c} + Y_c})/({a_{0c} +  b_{0c} + n_c}). 
\end{equation}
Given a reasonable $a>0$, when $Y_c/n_c \approx Y_{ch}/n_{ch}$, incorporating historical control reduces the sample size needed in the current study without introducing substantial bias. However, if  the response rates of the study control and that of historical control are substantially different, large $a$ leads to substantial bias \citep{2015_Power_Ibrahim}. In our proposed framework, we set $a = n_{ch, e}/n_{ch}$, where the maximum information borrowing is limited to equivalent $n_{ch, e}$ subjects among the total of $n_{ch}$ subjects borrowed. This will pose a global control of the influence from the historical study. In addition, in the proposed DPP framework, the operating characteristics of a design depend on all three borrowing parameters together. The simulation studies in Section \ref{sec.results} investigate their practical use.

%\subsection{How much should we borrow from historical study?}
%This is a critical question. Currently, we consider a naive approach: $n_{che} = n_c$; but what is the statistical justification from methodology perspective? What problems will happen if we borrow $n_{che} > n_c$?

  %%%%%%%%%%%%%%%%%%%%%%%%  
\subsection{Dynamic Borrowing} \label{sec.dynamicBorrowing}
One major component of the proposed Bayesian hybrid design framework involves determining the dynamic borrowing parameter $w_d$ as a measure of the similarity of $p_c$ between the current control and historical control. In this section, we consider several statistical methods for this purpose including empirical Bayes (EB), Bayesian $P$ (Bp), generalized Bhattacharyya coefficient (GBC), and Jensen-Shannon divergence (JSD) methods.

\vspace{1em}  
\textbf{Empirical Bayes}
 
The informative prior of $p_c$ given historical control before considering the global borrowing parameter $a$ is $p_c|Y_{ch}, a_{0c}, b_{0c} \sim Beta(a_{0c} + {w_d}Y_{ch}, b_{0c} + {w_d}(n_{ch}-Y_{ch}))$. One approach to determine $w_d$ is empirical Bayes, which is considered in the context of basket design by Zhou et al.\citep{2023_zhou_bayesian}. This method determines $w_d$ by maximizing the marginal likelihood of $Y_c$ after integrating out $p_c$ from the joint distribution of $(Y_c, p_c)$ given the historical data. 
  \begin{eqnarray}
          w_d &=& \argmax_{w_d\in [0, 1]}\left\{\int_0^1 p(Y_{c}|p_c) p(p_c|Y_{ch}) d p_c\right\}.\nonumber
       \end{eqnarray}
The integration can be expressed by closed-form of beta functions. The derivation is given in the supplementary materials. This approach tends to give a larger $w_d$ when the historical control and concurrent control are more similar in response rate and it appears a reasonable approach as a measure of similarity. 
       
Table \ref{tab:EB} shows the determined values of $w_d$ for various response rates in concurrent control, while the response rate in historical control is fixed at 0.30, under three different hyperpriors $Beta(0.001, 0.001)$, Jeffrey's prior $Beta(0.5, 0.5)$ and uniform prior $Beta(1,1)$. When the observed response rate in the concurrent control is the same as the historical control 0.3, i.e., the perfect scenario, the borrowing weight is maximized to 1.0. Due to the discrete nature of the binomial distribution, the weights $w_d$ are not exactly symmetric to the perfect scenario of $\hat{p}_c = 0.3$. 
       
\begin{table}
\centering
\caption{Dynamic weight $w_d$ in various scenarios of response rate (ORR) in concurrent control $\hat{p}_c$ among 40 subjects, while the historical control $\hat{p}_{ch} = 0.3$ among 200 subjects. {\revone The borrowing weight approaches to 1.0 when the difference between $p_c$ and $p_{ch}$ diminishes and decreases as the prior-data conflict increases.}}
    \begin{tabular}{|c|c|c|c|}
    \hline
    &\multicolumn{3}{|c|}{$w_d$ with hyperprior $Beta(a_{0c}, b_{0c})$}\\
    \cline{2-4}
    $\hat{p}_c$&$(0.001, 0.001)$ & $(0.5, 0.5)$ & $(1, 1)$\\
    \hline
    0.1&  0.020& 0.015 & 0.014\\
    0.2&  0.155& 0.181 & 0.232\\
    0.3&  1.000& 1.000& 1.000\\
    0.4&  0.308& 0.236 & 0.194\\
    0.5&  0.040& 0.031 & 0.026\\
    \hline
    \end{tabular}
    \label{tab:EB}
\end{table}

\vspace{1em}         
\textbf{Bayesian $P$}

The posterior distribution of response rate based on historical control $p_{ch} \sim Beta(a_{0c} + aY_{ch}, b_{0c} + a(n_{ch}-Y_{ch}))$ and based on concurrent control $p_{c}\sim Beta(a_{0c} + Y_{c}, b_{0c} + (n_{c}-Y_{c}))$. One can consider the heterogeneity measured based on the probability of response rate difference between these two posterior distributions with a tuning parameter $\eta > 0$,
    \begin{equation}
     w_d = \left[2\min (\xi_1, \xi_2)\right]^\eta,\nonumber
    \end{equation}
    where $\xi_1 =P(p_{c} \ge p_{ch})$ and $\xi_2 = P(p_{c} \le p_{ch})$. This can be interpreted as a 2-sided Bayesian $P$ value to a power of $\eta$.
    
 %%%%%%%%%%%%%%%%%%%%%%    
\vspace{1em}  
\textbf{Generalized Bhattacharyya Coefficient (BC)}

For simplicity, denote $f_c(x)$ and $f_{ch}(x)$ as the probability density functions of the response rate's posterior distributions under concurrent control $Beta(a_{0c} + Y_{c}, b_{0c} + (n_{c}-Y_{c}))$ and under historical control $Beta(a_{0c} + aY_{ch}, b_{0c} + a(n_{ch}-Y_{ch}))$ respectively. Bhattacharyya coefficient (BC) \citep{1943_Bhattacharyya} measures the amount of overlap between two statistical distributions with a value between 0 and 1. So it can be used to measure the similarity in our setting with $BC = \int_0^1 \sqrt{f_c(x) f_{ch}(x)} dx$, which can also be interpreted as the expectation of the density ratio:
    \begin{equation}
        BC = E_c\left[\left(\frac{f_{ch}(x)}{f_c(x)}\right)^{\frac{1}{2}}\right]
        = E_{ch}\left[\left(\frac{f_c(x)}{f_{ch}(x)}\right)^{\frac{1}{2}}\right],\nonumber
    \end{equation}
where $E_c$ and $E_{ch}$ denote the expectations with respect to $f_c(x)$ and $f_{ch}(x)$, respectively. We generalize the Bhattacharyya coefficient to facilitate the fine tuning of borrowing with $\theta$ and $\eta$ parameters below:
    \begin{eqnarray}
    w_d &=&\left\{\frac{1}{2}\left[\int_0^1 \left(\frac{f_{ch}(x)}{f_c(x)}\right)^{\theta}f_c(x)dx + \int_0^1 \left(\frac{f_{c}(x)}{f_{ch}(x)} \right)^{\theta}f_{ch}(x)dx\right]\right\}^\eta. \nonumber
  \end{eqnarray}
   When $f_c(x)$ and $f_{ch}(x)$ are identical, $w_d=1$. As a special case, when $\theta = \frac{1}{2}$ and $\eta=1$, this reduces to the Bhattacharyya coefficient. In addition, it can be shown that $w_d$ is symmetric for $\theta$ and $1-\theta$ when $\eta$ is fixed, so the effective range of $\theta$ is from 0 to 0.5. The dynamic borrowing parameter $w_d$ is a decreasing function of $\theta$ and $\eta$. 
   
   %\begin{figure}
   % \centering
   %	\includegraphics[height=3.5in]{wd_theta.png}
%	\caption{Dynamic borrowing using generalized BC method}
%	\label{fig.wd_theta}
 %   \end{figure}

\vspace{1em}
\textbf{Jensen-Shannon divergence (JSD) Method}

Fujikawa et al.\citep{2020_Fujikawa} consider a measure of similarity based on Jensen-Shannon divergence \citep{2004_Fuglede}. Let
    \begin{equation}
        w_d^* = 1-\frac{1}{2}\left[KL(f_c(x) \vert \frac{f_c(x)+f_{ch}(x)}{2})+KL(f_{ch}(x) \vert \frac{f_c(x)+f_{ch}(x)}{2})\right]\nonumber,
    \end{equation}
    where $KL(f_1(x)|f_2(x)) = \int_0^1 \log\left(\frac{f_1(x)}{f_2(x)}\right)f_1(x)dx$ is the Kullback-Leibler divergence between densities $f_1(x)$ and $f_2(x)$. 
    Then $w_d = (w_d^*)^\eta I(w_d^*>\tau^*)$. It is suggested to use $\eta=2$ by Fujikawa et al.\cite{2020_Fujikawa}. {\revone 
    $\tau^*$ is a threshold parameter. If the similarity is smaller than $\tau^*$, the two distributions, such as those of the study control and historical control, are treated as completely different. In the DPP framework, no borrowing from the historical control occurs when $w_d = 0$.
    } Since we already have a gated control parameter for the weight, we will consider $w_d = (w_d^*)^\eta$ and keeps $\eta$ as a tuning parameter.

%    \begin{figure}
%    \centering
% 	 \includegraphics[height=3.5in]{wd_JSD.png}
% 	 \caption{Dynamic borrowing using JSD method when $n_c=40$ and $p_c = 0.3$}
%	 \label{fig.wd_JSD}
%    \end{figure}
%\section{Methodology Questions}
%\subsection{Frameworks besides power prior}
%Need to consider several or at least 2 frameworks for modelling in design; to demonstrate the power prior is a reasonable choice. 

%\subsection{Success Boundary $\tau$}

%Need to understand the criteria for success. What would be the good recommendation of tau, 0.9 or 0.95. 

%If the actual sample size is different from the planned, we may need to calibrate tau to ensure an acceptable type I error. 

\begin{figure}
    \centering
    \includegraphics[width=\textwidth]{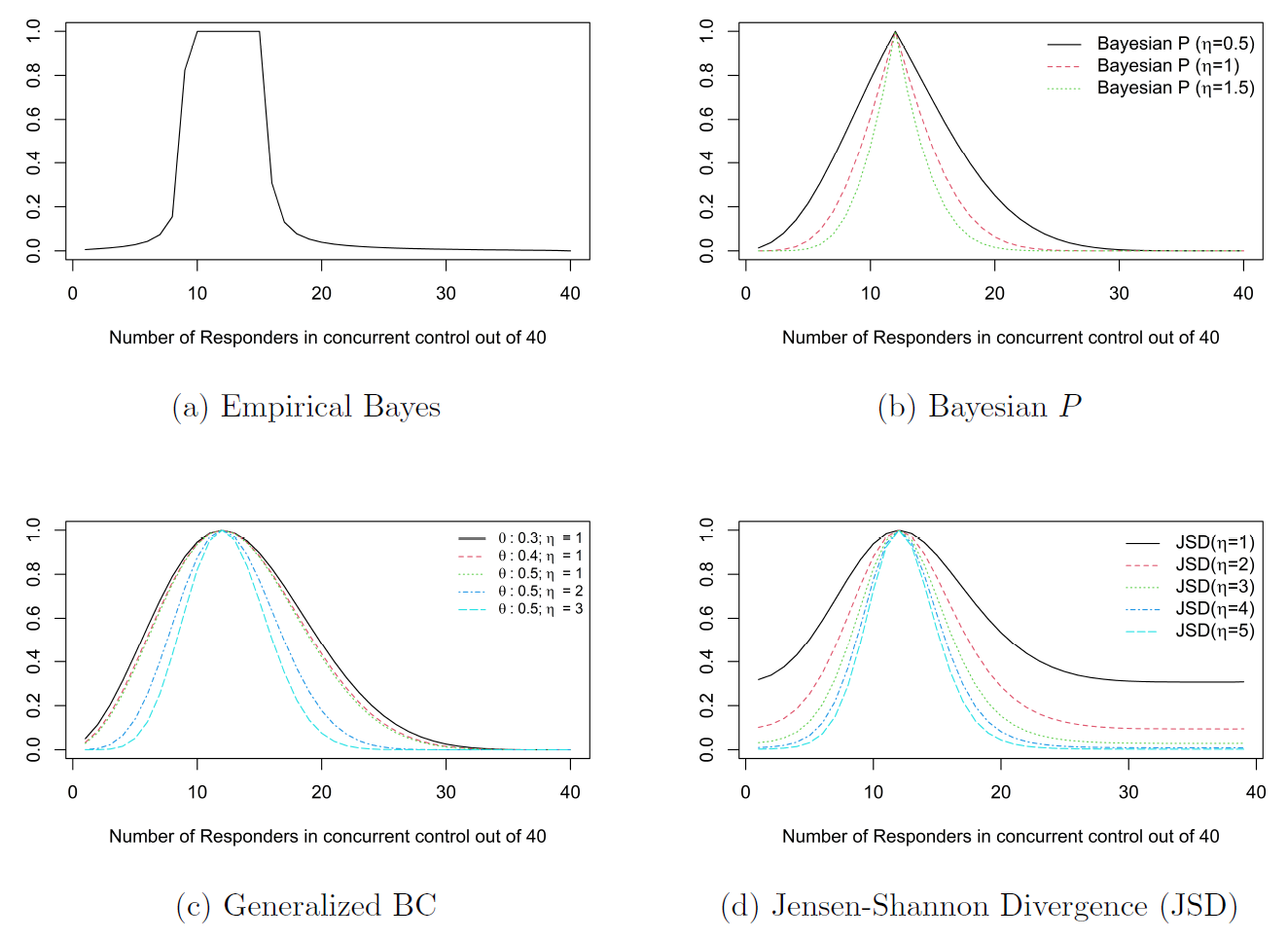}    
    \caption{Dynamic Borrowing Methods ($w_d$). {\revone 
    The borrowing weight approaches 1.0 as the difference between $p_c$ and $p_{ch}$ diminishes, and it decreases as the prior-data conflict increases. Different dynamic borrowing methods offer varying mechanisms for dynamic borrowing, and these can be optimized and determined through simulation by balancing levels of bias and power gain.
    } }
    \label{fig:wd}
\end{figure}

%\subsection{Impacts of hyper priors}
%This has been addressed for impacts on dynamic borrowing. But the impacts on type I error, power, and bias should also be assessed.
Figure \ref{fig:wd} illustrates the dynamic borrowing weight $w_d$ using the four methods in the case of $n_c=40$. When the observed response rate in the concurrent control is similar to the historical control, the empirical Bayes approach has the maximum borrowing with the weight approaching to 1.
\subsection{Further Considerations of Information Borrowing Features}
The proposed method provides flexible control of information borrowing from historical control. Power prior \citep{2015_Power_Ibrahim} can be considered as a special case of the proposed framework by setting the dynamic borrowing parameter $w_d=1$, the gated borrowing parameter $\Delta_{max}=\infty$ and the global borrowing parameter $a$ to a pre-determined value.
{\revone 
A dynamic power prior for continuous endpoints in noninferiority trials\citep{2024_A_Mariani} was proposed, where the prior includes a global borrowing parameter and a dynamic borrowing parameter determined by the Hellinger distance. However, this framework does not include a gated borrowing parameter to guard against large discrepancies between the current control and the historical control, and it requires an MCMC algorithm, which may be time-consuming for both simulation and calibration.
}

Meta-analytic predictive (MAP) prior  \citep{2010_Neuenschwander_Summarizing} uses a hierarchical model to borrow information from multiple external control groups. The model is similar to mixed effect models where the randomized controlled trial and external control groups are subgroups/clusters. The quantities of interest of all subgroups, e.g., mean or proportions, are assumed to follow a distribution and are shrunk towards a common parameter. Robust meta-analytic predictive prior (rMAP) \citep{2014_rMAP_Schmidli} improves MAP by mixing MAP with a non-informative prior distribution to better address significant prior-data discrepancy. MCMC algorithms, kernel-density estimates and mixture modeling methods are used to elicit the informative prior distribution. 
One challenge of rMAP is the choice of mixture weight of MAP relative to the non-informative prior. The self-adapting mixture (SAM) prior \citep{2023_SAM_Yang} dynamically determines the weight through a likelihood ratio test or posterior probability ratio, quantifying the extent of prior data discrepancy. The MAP prior can serve as the informative prior within the SAM prior. In cases of substantial discrepancy, the weight for the non-informative component derived from concurrent data tends to be high, potentially causing the SAM prior to degenerate to a non-informative prior. This parallels the concept of dynamic borrowing within the Bayesian DPP framework. One advantage of power prior is the ease of interpretation of the effective sample size borrowed from the historical control. According to Equation (\ref{eq.posterior}), the effective sample size of borrowing is $n_{ch}w$. As shown in Equation (\ref{eq:w}), $w$ is a function of $Y_c$ which follows a binomial distribution. We define the expected effective sample size (EESS) to evaluate the expected amount of borrowing as 
$$EESS = n_{ch,e}\sum_{Y_c=0}^{n_c} w P(Y_c|n_c,p_c).$$ 
A statistical design with higher $EESS$ indicates more information borrowing. 
In addition to the dynamic borrowing feature, the proposed DPP framework also incorporates a maximum allowable amount of borrowing ($n_{ch, e}$) and a tolerance level of discrepancy ($\Delta_{max}$). These aspects are practically important for data interpretation, as they help strike a balance between leveraging information from historical studies and emphasizing the current study's findings. 
%In Bayesian modeling, hyperpriors are necessary components. In our investigation, we examined the effects of vague hyperpriors, details provided in Supplementary Materials. Overall, the impacts on dynamic borrowing appear to be quite minimal when considering conjugate hyperpriors such as $Beta(0.001, 0.001)$, Jeffrey's prior $Beta(0.5, 0.5)$, and a uniform prior $Beta(1, 1)$.

\subsection{Calibration and Optimization}

Type I error rate is the probability of claiming statistical significance while assuming null hypothesis is true, i.e., the response rates of experimental treatment and concurrent control in current study are equal.
In the Bayesian Hybrid Design in DPP framework, the study is claimed statistically significant if 
\begin{equation}\label{eq.sig}
  P(p_t > p_c|\text{hybrid data}) > \tau, 
\end{equation}
where $\tau$ is a predefined threshold obtained by calibration using simulations under null hypothesis to control a prespecified type I error rate $\alpha$.
When $\tau$ is calibrated assuming equal response rates in the concurrent control and historical control ($p_c=p_{ch}$), type I error will be likely inflated if the concurrent control's response rate is greater than that of the historical control. 
We can explore various design options of the proposed Bayesian DPP framework with borrowing from external control and strategically optimize the design to achieve acceptable type I error rate and while avoiding unacceptable bias. The optimization can be performed by considering a suitable dynamic borrowing method and appropriate amount of information borrowed from external control.

In addition to the traditional requirements of type I error and power in conventional randomized controlled trials, Bayesian hybrid designs should also consider the tolerable level of influence from historical studies for evaluating design performance. The influence from historical control always increases with more information borrowed ($n_{ch,e}$) when concurrent control and historical control differ in observed response rate. One approach to evaluate the influence is to calculate the posterior mean difference $d=\mu - \tilde{\mu}$ between the point estimates with and without incorporating historical control (Eq. (\ref{eq.post.mean}) and Eq. (\ref{eq.post.mean.nohybrid})). The conditional expectation of $E(d|\text{historical data})$, termed mean posterior mean difference (PMD), or the probability of absolute PMD exceeding certain threshold $\epsilon>0$, i.e., $\xi(\epsilon) = P(|d| > \epsilon | \text{historical data})$, can be used to assess the influence of historical control. The expectation and the probability are calculated by integrating out $Y_c$, and both quantities can be estimated via Monte Carlo simulations. The design optimization should achieve adequate power, maintain type I error, and constrain the influence of historical data within a certain tolerable level $E(d|\text{historical data})\le d^*$ or $\xi(\epsilon) \le \xi^*$. The optimization is illustrated in a case study in Section \ref{sec.example}.

\section{Results} \label{sec.results}
\subsection{Simulations}
To demonstrate the borrowing mechanism in the proposed DPP framework, we performed a simulation study and explore the performance. We consider a hybrid study design in which 90 subjects are equally randomized into two groups: an experimental treatment group (comprising a novel agent combined with standard of care (SOC)) and a SOC-only treatment group ($n_c=n_t=45$). Furthermore, the hybrid control incorporates subjects from a previously completed large randomized phase 3 study, which included 180 eligible patients treated with SOC ($n_{ch}=180$). Additionally, the observed response rate of the SOC arm in the historical study is 0.3. For the purpose of power calculation, we assume a 20\% increase in the response rate for the experimental treatment.

To gauge the effectiveness of the DPP borrowing mechanism, three methods are examined in this simulation study:
(1) DPP with $\Delta_{max}=0.1$ and dynamic borrowing utilizing the empirical Bayes approach.
(2) DPP with $\Delta_{max}=\infty$, indicating no gated borrowing control, using the empirical Bayes approach.
(3) Fixed power prior: This method excludes dynamic borrowing and gated borrowing ($w_d=1$ and $\Delta_{max}=\infty$), with the fixed power parameter determined by $a$. This can be seen as a special case of the proposed DPP framework.
We evaluate a total of 28 scenarios including all combinations of 4 values for the global borrowing parameter $a = 1/4, 1/2, 3/4, 1$, corresponding to the equivalent numbers of subjects borrowed from the historical study $n_{ch,e} = 45, 90, 135, 180$, and 7 values for the concurrent control response rate $p_{c}=0.15, 0.20, 0.25, 0.30, 0.35, 0.40, 0.45$. In each scenario, hyperpriors for both control and experimental response rates are set as $Beta(0.001, 0.001)$, and 100,000 simulated trials are conducted. 

{\revone Additionally, $\tau$ is obtained by calibration to maintain a type I error rate of 0.1 for $p_c=p_{ch}=0.3$. More specifically, we simulated 100,000 data sets with $p_c=p_{ch}=0.3$ and other settings indicated above and used the 90\% percentile of the posterior probability in Equation (\ref{eq.sig}) as $\tau$. The concurrent control might have a different $p_c$ due to temporal effect and heterogeneity. The values of $p_{c}$ other than 0.3 are considered to assess the sensitivity of the study design.}

The estimated type I error, power, mean, and standard deviation of posterior mean difference (PMD) are summarized in Table \ref{tab:Sim1Res}. {\revone When $p_c=p_{ch}=0.3$, the type-I error rates for DPP, SAMprior and rMAP are less than and close to $0.1$, indicating effective calibration of the type-I error rates.} In situations where the response rate of the concurrent control exceeds that of the external control ($p_{c} > 0.3$), the fixed power prior approach without gated borrowing control or dynamic borrowing control (Method (3)) may lead to considerable inflation in type I error. After integrating dynamic borrowing control (Method (2) with $\Delta_{max}=\infty$) can notably mitigate this inflation, as it proportionally reduces the amount of borrowed information from the external control based on the disparity in response rates between the external and concurrent controls. Moreover, the inclusion of both gated borrowing control and dynamic borrowing control (Method (1) with $\Delta_{max}=0.1$) further diminishes type I error inflation in such scenarios. The gated borrowing parameter specifically diminishes the amount of borrowed information from the external control, abstaining from borrowing altogether when the response difference exceeds 0.1. In situations where the response rate of the concurrent control lower than that of the external control ($p_{c} < 0.3$), borrowing from the historical study does not lead to inflation in type I error or power. This is attributed to the effective control of borrowing by the DPP methods. Notably, the DPP methods also sustain power levels due to their adaptive borrowing mechanism. The influence of external data on the posterior mean is one important consideration in hybrid design. By using the proposed borrowing control mechanism in the DPP framework, the influence is significantly mitigated.

{\revone We conduct additional simulation studies to demonstrate the impact of the gated control parameter. We repeat the simulations using DPP with $\delta_{max}$ set to 0.05 and 0.15. The results are presented in Supplementary Table S1. When $p_c < 0.3$, the different $\delta_{max}$ values lead to negligible changes in type I error and power. However, when $p_c > 0.3$, the smaller the $\delta_{max}$, the smaller the type I error and power. On average, less information is borrowed from historical data because borrowing is less likely to occur when the difference between $\hat{p}_c$ and $\hat{p}_{ch}$ exceeds $\delta_{max}$.
}

We further illustrate the proposed DPP framework by comparing it to rMAP \citep{2014_rMAP_Schmidli} and SAMprior \citep{2023_SAM_Yang} methods using additional dynamic borrowing methods within the same settings as described above. {\revone It was demonstrated that SAMprior yields superior performance compared to non-informative prior, power prior, and commensurate prior.\citep{2023_SAM_Yang} The rMAP\citep{2014_rMAP_Schmidli} includes an additional vague component in the mixture distribution to address potential prior-data conflict compared to the MAP. 
An alternative approach is to assign a prior distribution to the borrowing weight. In a previous study\citep{2017_Gravestock_Adaptive}, the power prior approach with the overall borrowing weight modeled by a prior distribution was compared to that by an Empirical Bayes method. Compared to the Empirical Bayes method, this approach leads to less borrowing when $\hat{p}_c$ and $\hat{p}_{ch}$ are similar, while resulting in more borrowing when $\hat{p}_c$ and $\hat{p}_{ch}$ are very different. Essentially, this approach represents an alternative way, based on the joint posterior distribution of the borrowing weight $w$ and response rate $p_c$, to determine the amount of information borrowing according to $\hat{p}_c - \hat{p}_{ch}$, similar to those shown in Figure \ref{fig:wd}. The proposed DPP framework allows more flexibility in determining such a relation. Moreover, the marginal posterior distributions of the approach that assigns prior to the weight parameter are nonstandard or without a closed form\citep{2017_Gravestock_Adaptive}. Consequently, the computation is likely to be substantially more time-consuming than with the proposed DPP framework.

} 

{\revone We use the default parameters in the SAMprior R package, e.g., clinical significant different = 0.1, for the analyses with SAM prior and rMAP. The gated borrowing parameter in DPP agrees with the clinical significant different in SAMprior, which facilitates a fair comparison.
The significance thresholds for SAM prior and rMAP are also calibrated using data generated from $p_c = p_t = \hat{p}_{ch}$ to achieve type I error rate of 0.1.}
 
Within the DPP framework, we consider the following dynamic borrowing methods: empirical Bayes (EB), Bayesian $P$ (Bp) with $\eta=1$, generalized BC method with $\theta=0.5$ and $\eta=1$, and JSD with $\eta=2$. The historical response rate ($p_{ch}$) remains fixed at 0.3 for all scenarios. We evaluate the performance of these methods based on type I error, power, mean and standard deviation of posterior mean difference (PMD), and computational time. Assuming a 20\% increase in response rate in the experimental treatment arm when estimating power. Overall, the type I errors for the four dynamic borrowing methods within the proposed DPP framework are generally comparable to SAM and rMAP methods. SAM achieves a higher power when the current control has the same response rate as the historical control, however its mean PMD is much larger than the Bayesian DPP methods when a discrepancy exists between historical and concurrent controls. This demonstrates the importance of having global control parameter over borrowing to restrict the influence from historical control. 

Furthermore, all dynamic borrowing methods in the proposed DPP framework are based on closed-form solutions, making them more efficient than methods based on Markov Chain Monte Carlo methods. For instance, the computational time required to run 100,000 simulated trials using SAM or rMAP methods is approximately 70 times slower than the DPP methods. This efficiency is particularly crucial when exploring numerous design settings.

\begin{landscape}
    
\begin{table}[ht]
\begin{center}
\caption{Dynamic Borrowing Mechanism in the Proposed DPP Framework: Type I Error, Power, Mean and Standard Deviation of Posterior Mean Difference (PMD)}\label{tab:Sim1Res}
{\footnotesize
\begin{tabular}{l|r|cccrrrrrrrrrrr}
\hline
  && \multicolumn{3}{c}{\makecell{Dynamic power prior \\ $\Delta_{max}=0.1$}}& \multicolumn{3}{c}{\makecell{Dynamic power prior\\ $\Delta_{max}=\infty$}}& \multicolumn{3}{c}{\makecell{Fixed power prior \\ 
  $\Delta_{max}=\infty$ and $w_d=1$}}& Bp& GBC& JSD& SAM&rMAP\\ \cline{3-11}
  &$p_{c} \backslash n_{ch,e}$& 
  \makecell{45}& \makecell{90}&\makecell{180}&
  \makecell{45}& \makecell{90}&\makecell{180}& 
  \makecell{45}& \makecell{90}&\makecell{180}&
  \makecell{45}& 
  \makecell{45} & \makecell{45}& ---& ---\\ 
 \hline
 \multirow{7}*{\makecell{Type I\\error\\rate}}& 0.15& 0.103 & 0.118 & 0.103 & 0.101 & 0.095 & 0.076 & 0.004 & 0.001 & 0.000 
& 0.116 & 0.119 & 0.119 
& 0.084 &0.062 
\\ 
&   0.20& 0.104 & 0.106 & 0.102 & 0.095 & 0.084 & 0.072 & 0.019 & 0.008 & 
 0.005 
& 0.106 & 0.113 & 0.113 
& 0.069 &0.065 
\\ 
&   0.25& 0.096 & 0.090 & 0.087 & 0.090 & 0.080 & 0.070 & 0.047 & 0.035 &  0.027 
& 0.093 & 0.098 & 0.098 
& 0.073 &0.078 
\\ 
&   0.30& 0.099 & 0.099 & 0.098 & 0.099 & 0.097 & 0.098 & 0.097 & 0.099 & 
  0.099 
& 0.100 & 0.099 & 0.099 
& 0.099 &0.099 
\\ 
&   0.35& 0.127 & 0.146 & 0.160 & 0.134 & 0.149 & 0.175 & 0.172 & 0.202 &  0.238 
& 0.128 & 0.133 & 0.133 
& 0.131 &0.119 
\\ 
&   0.40& 0.150 & 0.187 & 0.205 & 0.170 & 0.202 & 0.246 & 0.269 & 0.337 & 
  0.435 
& 0.147 & 0.165 & 0.165 
& 0.157 &0.140 
\\ 
&   0.45& 0.133 & 0.170 & 0.165 & 0.165 & 0.202 & 0.248 & 0.377 & 0.498 &  0.644 
& 0.143 & 0.157 & 0.157 
& 0.168 &0.151 
\\ 
   \hline
 \multirow{7}*{Power}& 0.15
& 0.821 & 0.810 & 0.803 & 0.811 & 0.793 & 0.769 & 0.654 & 0.541 & 
  0.454 
& 0.814 & 0.822 & 0.822 
& 0.773 &0.783 
\\ 
&   0.20
& 0.782 & 0.772 & 0.758 & 0.780 & 0.769 & 0.751 & 0.742 & 0.702 &  0.657 
& 0.781 & 0.776 & 0.776 
& 0.776 &0.784 
\\ 
&   0.25
& 0.794 & 0.804 & 0.799 & 0.799 & 0.806 & 0.814 & 0.825 & 0.829 & 
  0.820 
& 0.799 & 0.796 & 0.796 
& 0.806 &0.794 
\\ 
&   0.30
& 0.811 & 0.842 & 0.843 & 0.833 & 0.857 & 0.881 & 0.889 & 0.910 &  0.925 
& 0.815 & 0.827 & 0.827 
& 0.826 &0.806 
\\ 
&   0.35
& 0.779 & 0.823 & 0.799 & 0.824 & 0.859 & 0.889 & 0.935 & 0.958 & 
  0.975 
& 0.803 & 0.814 & 0.814 
& 0.833 &0.813 
\\ 
&   0.40
& 0.733 & 0.791 & 0.738 & 0.795 & 0.830 & 0.863 & 0.966 & 0.984 &  0.994 
& 0.785 & 0.789 & 0.789 
& 0.834 &0.799 
\\ 
&   0.45& 0.720 & 0.779 & 0.720 & 0.781 & 0.800 & 0.845 & 0.984 & 0.995 & 
  0.999 
& 0.778 & 0.779 & 0.779 
& 0.836 &0.789 
\\
   \hline
   \multirow{7}*{\makecell{Mean \\ PMD}}& 0.15
& 0.004 & 0.006 & 0.009 & 0.009 & 0.015 & 0.025 & 0.075 & 0.100 &  0.120 
& 0.004 & 0.008 & 0.008 
& 0.026 &0.024 
\\ 
&   0.20
& 0.009 & 0.014 & 0.019 & 0.012 & 0.020 & 0.029 & 0.050 & 0.067 & 
  0.080 
& 0.009 & 0.014 & 0.015 
& 0.024 &0.021 
\\ 
&   0.25
& 0.008 & 0.012 & 0.016 & 0.009 & 0.014 & 0.020 & 0.025 & 0.033 & 0.040 
& 0.008 & 0.012 & 0.013 
& 0.014 &0.013 
\\ 
&   0.30
& 0.000 & 0.001 & 0.002 & -0.001 & -0.001 & -0.001 & 0.000 & 0.000 & 
 0.000 
& 0.001 & 0.002 & 0.003 
& 0.002 &0.004 
\\ 
&   0.35
& -0.007 & -0.009 & -0.011 & -0.011 & -0.017 & -0.023 & -0.025 & -0.033 & -0.040 
& -0.004 & -0.006 & -0.006 
& -0.010 &-0.004 
\\ 
&   0.40
& -0.008 & -0.011 & -0.013 & -0.016 & -0.024 & -0.035 & -0.050 & -0.067 & 
 -0.080 
& -0.005 & -0.007 & -0.007 
& -0.020 &-0.010 
\\ 
&   0.45& -0.005 & -0.006 & -0.008 & -0.014 & -0.023 & -0.036 & -0.075 & -0.100 & -0.120 
& -0.003 & -0.005 & -0.005 & -0.024 &-0.013 
\\ 
   \hline
 \multirow{7}*{\makecell{sd(PMD)}}& 0.15
& 0.009 & 0.013 & 0.018 & 0.007 & 0.009 & 0.011 & 0.027 & 0.035 & 
 0.043 
& 0.008 & 0.015 & 0.015 
& 0.008 &0.004 
\\ 
&   0.20
& 0.012 & 0.017 & 0.023 & 0.010 & 0.014 & 0.017 & 0.030 & 0.040 & 0.048 
& 0.011 & 0.018 & 0.018 
& 0.011 &0.008 
\\ 
&   0.25
& 0.016 & 0.023 & 0.029 & 0.016 & 0.023 & 0.029 & 0.032 & 0.043 & 
 0.052 
& 0.013 & 0.020 & 0.020 
& 0.016 &0.012 
\\ 
&   0.30
& 0.019 & 0.026 & 0.033 & 0.020 & 0.028 & 0.036 & 0.034 & 0.045 & 0.055 
& 0.015 & 0.021 & 0.021 
& 0.018 &0.013 
\\ 
&   0.35
& 0.018 & 0.024 & 0.030 & 0.018 & 0.025 & 0.032 & 0.035 & 0.047 & 
 0.057 
& 0.013 & 0.018 & 0.019 
& 0.017 &0.011 
\\ 
&   0.40
& 0.015 & 0.021 & 0.025 & 0.013 & 0.018 & 0.022 & 0.036 & 0.049 & 0.058 
& 0.010 & 0.015 & 0.015 
& 0.013 &0.008 
\\ 
&   0.45& 0.012 & 0.016 & 0.019 & 0.010 & 0.013 & 0.015 & 0.037 & 0.049 & 0.059 & 0.008 & 0.011 & 0.011 & 0.008 &0.005 \\ 

   \hline
\end{tabular}
} 
\end{center}
{\footnotesize Note: {\revone For all scenarios, $\tau$ is calibrated for $p_t=p_c=p_{ch}=0.3$.} For both dynamic power prior methods, empirical Bayes approach is used for dynamic borrowing. PMD: posterior mean difference in response rate. {\revone Power is calculated assuming 20\% improvement in response rate compared to $p_c$, i.e., $p_t=0.5$.} Bp: Bayesian $P$ with $\eta=1$; GBC: Generalized BC with $\theta=0.5, \eta=1$; JSD: $\eta=2$.}
\end{table}
\end{landscape}

%To assess the sensitivity to the gated borrowing parameter ($\Delta_{max}$) given different dynamic borrowing method for the proposed framework, we repeated the analyses with larger and smaller gated borrowing parameter ($\Delta_{max}=0.15, 0.05$). The results are shown in Supplementary Table S2 and S3. 

%\begin{table}[ht]
%\begin{center}
%\caption{\textbf{DELETE THIS TABLE}. Computational time (in seconds) for 100,000 simulated trials}\label{tab:time}
%\begin{tabular}{|c|c|c|c|c|c|c|c|c|c|c|c|}\hline
% $n_t$ & $n_c$ & $n_{ch,e}$& $p_t$& $p_c$& $p_{ch}$ & EB&  Bp& GBC & JSD & SAM & rMAP\\ 
%  \hline
% 40.0 & 40.0 & 40.0 & 0.4 & 0.2 & 0.3 & 34.4 & 63.3 & 44.5 & 46.2 & 1931.5 & 1240.9 \\ 
%   40.0 & 40.0 & 40.0 & 0.5 & 0.3 & 0.3 & 33.3 & 60.8 & 43.5 & 43.7 & 1934.5 & 1255.8 \\ 
%   40.0 & 40.0 & 40.0 & 0.6 & 0.3 & 0.3 & 32.4 & 57.3 & 41.4 & 41.3 & 1911.5 & 1238.6 \\ 
%   45.0 & 45.0 & 45.0 & 0.4 & 0.2 & 0.3 & 33.5 & 63.5 & 47.4 & 48.0 & 1917.2 & 1243.3 \\ 
%   45.0 & 45.0 & 45.0 & 0.5 & 0.3 & 0.3 & 33.8 & 61.8 & 44.4 & 46.8 & 1919.4 & 1267.9 \\ 
%   45.0 & 45.0 & 45.0 & 0.6 & 0.3 & 0.3 & 32.4 & 58.6 & 43.0 & 44.5 & 1927.1 & 1262.0 \\ 
%   \hline
%\end{tabular}
%\end{center}
%\end{table}
  
\subsection{Example} \label{sec.example}
Pembrolizumab has been approved as the first-line treatment of patients with non-small cell lung cancer (NSCLC) expressing PD-L1 (Tumor Proportion Score (TPS) $\ge 1\%$) \citep{2021_FDA_HIGHLIGHTS}. According to the clinical trial KEYNOTE-042 \citep{2019_Mok_Pembrolizumab}, the estimated response rate is 27\% based on 637 subjects treated with pembrolizumab monotherapy. The rich resource of high-quality data provided by the historical study offers valuable opportunities for further clinical development, particularly in combination therapy with novel agents. Suppose we would like to design an early phase exploratory hybrid study for a combination therapy of a novel agent plus pembrolizumab and target at 20\% improvement of response rate. A traditional frequentist design of a randomized study with 80\% power and 10\% one-sided type I error will require about 135 subjects in 2:1 randomization ratio using Fisher's exact test, which can be challenging in feasibility at early stage of clinical development in cancer drugs \citep{2022_FDA_FIH}. A hybrid design in the proposed DPP framework can be used to improve statistical power by borrowing information. In order to ensure subjects are comparable to the current study, it is essential to access to the subject level data in the historical study to select eligible subjects and control confounding by known prognostics factors.  

As illustration,  two dynamic borrowing methods are used including empirical Bayes and Bayesian $P$ with $\eta=1$. We target at 80\% power and one-sided type I error 0.1. Table \ref{tab:example} shows the operating characteristics of both design methods. Compared to the traditional randomized controlled trial design, the hybrid design options have significant power increase and results in smaller sample size. Regarding the amount of information borrowing ($n_{ch, e}$), three scenarios are considered: $n_c$, $1.5n_c$, $2n_c$. For all scenarios, the calibration is performed for $p_c=p_{ch}=0.27$, as a result some type I error inflation is expected when borrowing from historical control ($p_{ch}$) that has a relatively smaller response rate. To assess the degree of type I error inflation, a range of $\pm 10\%$ discrepancy is considered in Table \ref{tab:example}, which represents a relatively conservative planning and the current study response rate is less likely to have a discrepancy more than 10\%. For each scenario, the minimum sample size that has at least 80\% power is summarized in Table \ref{tab:example}. We can observe that the current study sample size tends to decrease when borrowing more from historical control.
Regarding the amount of borrowing relative to concurrent control, the design with more borrowing is generally more favorable provided the mean (PMD) is tolerable.
Regarding comparison of different dynamic borrowing methods, in this example, Bayesian $P$ method produces a smaller average of absolute mean (PMD) than the empirical Bayes method. However, its power is also slightly lower with the same sample size of 56:28 than empirical Bayes method. 

\begin{figure}
\centering
\includegraphics[height=5in]{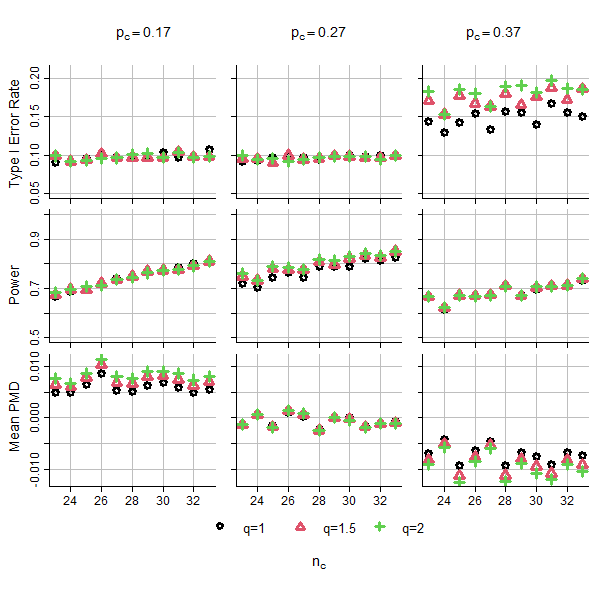}
\caption{Type I error rate, power, and mean PMD using empirical Bayes dynamic borrowing method. Calibration is performed assuming the concurrent control and historical control have the same response rate $(p_c=p_{ch}=0.27)$ for one-sided type I error of 0.1. Maximum amount of borrowing $n_{ch,e}=qn_c$.}
\label{fig:example}
\end{figure}

To further illustrate the considerations of sample size and study design operating characteristics, Figure \ref{fig:example} displays the power, type I error and mean (PMD) across a range of sample size of $n_c$ using empirical Bayes method. Figure S1 for Bayesian $P$ method is included in the supplementary materials. When the concurrent control and historical control have different response rates, the absolute mean PMD increases with more borrowing from historical control. If we would like to limit the absolute mean PMD within 1\% when the discrepancy is $\pm 10\%$, then the two design options are empirical Bayes approach with maximum borrowing of $n_{ch,e}=31$ and Bayesian $p$ approach with maximum borrowing of $n_{ch,e}=45$, where the empirical Bayes approach is slightly more powerful than the Bayesian $p$ approach in this case (0.822 vs 0.810) with slight increase of sample size from 90 to 93. 

{\revone 
When $p_{c}=0.37$, type I error rates are inflated because $p_{ch} < p_c$, and the posterior distribution of $p_c$ is pooled towards $p_{ch}$ due to borrowing information from the historical data. However, it is observed that the power with $p_{c}=0.37$ is smaller than that with $p_{c}=0.27$ and $0.17$, which seems to contradict the pooling effect mentioned above. Further investigation reveals the main reason that $p_{c}=0.37$ leads to a larger variance in the binomial distribution compared to $p_{c}=0.17$. Thus, when there is no borrowing from historical data, the power with $p_{c}=0.37$ is smaller than that with $p_{c}=0.17$, given the same sample size. We confirm this reasoning by repeating the analysis for Figure \ref{fig:example} without borrowing from historical data, with the results shown in Supplementary Figure S2. By comparing Figure \ref{fig:example} and Supplementary Figure S2, we demonstrate that the smaller power with historical borrowing for $p_c=0.37$ compared to $p_c=0.17$ is due to the inherently smaller power for $p_c=0.37$ when no borrowing from historical data occurs. The improvement in power due to borrowing from historical data is greater for $p_c=0.37$ than that for $p_c=0.17$ because $p_{ch} < p_c$ and $p_c$ is pulled towards $p_{ch}$. Nevertheless, the improvement in both cases is smaller than that for $p_c = 0.27$, owing to the gated control parameter in the proposed DPP framework.
}

We also assessed the situation where a significant number of subjects from the historical study are ineligible for inclusion in the analysis, i.e., $n_{ch}<637$. If the response rate remains the same based on the selected eligible subjects and the amount of borrowing $n_{ch, e}$ remains constant, the impact on the posterior distribution (\ref{eq.posterior}) is minimal. This is because $wY_{ch} \approx n_{ch, e}p_{ch} w_d I(|\hat{p}_c-\hat{p}_{ch}|< \Delta_{max})$, and $w_d$, as the measure of similarity, does not change significantly when the response rate remains constant. 

Regarding the randomization ratio in the concurrent study, e.g., 1:1 versus 2:1,  in general, 2:1 is more advantageous than 1:1 because the historical study can supplement the standard-of-care (SOC) treatment group. However, practical considerations such as interim analysis decision making and biomarker development considerations may favor equal allocation. Additionally, concerns about significant heterogeneity in efficacy due to temporal effects, population heterogeneity, or variations in medical practices can limit the benefits of borrowing. These factors collectively influence the choice of design options according to specific circumstances of each study. 

\begin{table}
\centering
\caption{Hybrid design for a novel combination therapy with borrowing from a historical trial KEYNOTE-042 of monotherapy in NSCLC  using empirical Bayes and Bayesian $P$ methods}
\label{tab:example}
\centering
\begin{tabular}[t]{ccccccccccc}
\hline
\makecell{Dynamic \\Method}  &  $n_t:n_c:n_{ch,e}$&$n_t$ & $n_c$ & $n_{ch,e}$ &EESS& $p_c$ & $p_t$ & Type I Error & Power &\makecell{Mean \\PMD(\%)}\\
\hline
\multirow{9}{*}{\makecell{Empirical\\ Bayes}}&  2:1:1&62 & 31 & 31  &11.01& 0.17 & 0.37 & 0.0971 & 0.785 
& 0.65 \\
&  &62 & 31 & 31  &22.47& 0.27 & 0.47 & 0.0975 & 0.822 
& -0.13 \\
&  &62 & 31 & 31  &15.27& 0.37 & 0.57 & 0.1666 & 0.708 
& -0.90 
\\
&  2:1:1.5
&56 & 28 & 42  &14.92& 0.17 & 0.37 & 0.1011 & 0.744 
& 0.70 
\\
&  
&56 & 28 & 42  &29.88& 0.27 & 0.47 & 0.0964 & 0.817 
& -0.22 
\\
&  
&56 & 28 & 42  &21.26& 0.37 & 0.57 & 0.1893 & 0.710 
& -1.12 
\\
&  2:1:2&56 & 28 & 56  &19.90& 0.17 & 0.37 & 0.1011 & 0.744 
& 0.82 
\\
&  &56 & 28 & 56  &39.84& 0.27 & 0.47 & 0.0964 & 0.817 
& -0.22 
\\
&  &56 & 28 & 56  &28.34& 0.37 & 0.57 & 0.1893 & 0.710 & -1.24 
\\
\hline
\multirow{9}{*}{\makecell{Bayesian $p$\\ ($\eta=1$)}}&  2:1:1 &64 & 32 & 32  &9.19& 0.17 & 0.37 & 0.1055 & 0.789 
& 0.67 
\\
& 
&64 & 32 & 32  &17.31& 0.27 & 0.47 & 0.0957 & 0.805 
& 0.10 
\\
&  &64 & 32 & 32  &9.93& 0.37 & 0.57 & 0.1463 & 0.710 
& -0.47 
\\
&  2:1:1.5
&60 & 30 & 45  &11.62& 0.17 & 0.37 & 0.0966 & 0.774 
& 0.58 
\\
&  &60 & 30 & 45  &23.75& 0.27 & 0.47 & 0.0971 & 0.810 
& -0.16 
\\
&  &60 & 30 & 45  &15.46& 0.37 & 0.57 & 0.1562 & 0.705 
& -0.95 
\\
&  2:1:2&56 & 28 & 56  &16.19& 0.17 & 0.37 & 0.1029 & 0.752 
& 1.08 
\\

&  &56 & 28 & 56  &28.97& 0.27 & 0.47 & 0.0994 & 0.806 
& 0.08 
\\
&  &56 & 28 & 56  &18.46& 0.37 & 0.57 & 0.1734 & 0.708 & -0.88 
\\
\hline
\end{tabular}
{\footnotesize Notes: Calculations are based on 100,000 simulated trials for each scenario. Calibration is performed assuming the concurrent control and historical control have the same response rate $(p_c=p_{ch}=0.27)$ for one-sided type I error rate of 0.1. EESS: Expected effective sample size borrowed from historical study.}
%Maximum amount of borrowing $n_{ch,e}=qn_c$. 
\end{table}

%Figure \ref{fig:power.nche} displays the power by maximum amount of borrowing ($n_{ch, e}$) in study design setting: %$p_t=0.47$, $n_t=60$, $p_c=0.27$, $n_c=30$, $p_{ch}=0.27$, $n_{ch}=637$, %$\alpha=0.1$,$a_{0c}=0.001$,$b_{0c}=0.001$,$a_{0t}=0.001$,$b_{0t}=0.001$,$\Delta_{max}=0.1$.
%\begin{figure}
%    \centering
%    \includegraphics[width=0.7\linewidth]{power_nche.png}
%    \caption{Power of Bayesian DPP by Maximum Amount of Borrowing $n_{ch, e}$}
%    \label{fig:power.nche}
%\end{figure}

\section{Conclusions and Discussions} \label{sec.discussion}
The proposed Dynamic Power Prior (DPP) framework was inspired by a practical case in combination therapy development, aiming to maximize the utilization of high-quality clinical data from previous trials. It is tailored for designing exploratory early-phase cancer trials that leverage external information to enhance power and increase trial efficiency. The flexible DPP framework provides a clear interpretation of the borrowing mechanism and enables customized control of borrowing to fit specific study settings through three key components aimed at mitigating risks of potential prior data discrepancy: global control, gated control, and dynamic control. Simulation studies have demonstrated its utility. The explicit interpretation of the DPP framework is particularly valuable in practice for facilitating cross-functional discussions to optimize a study design. The global control parameter effectively regulates the influence of historical control, regardless of the sample size in the historical control group. This aspect is very important at the study design stage, especially when determining the actual number of historical control subjects by using propensity score matching methods. In contrast, the SAM prior lacks maneuverability in global borrowing control, making it challenging to manage the influence of historical control at time of study design. 
Additionally, the proposed Bayesian DPP framework demonstrates significantly enhanced computational efficiency. The DPP framework offers a wide array of options for constructing a study design. Extensive simulations are needed to optimize a specific study design, as demonstrated in the example. It is essential to thoroughly examine various dynamic borrowing methods and amounts of borrowing to quantify their impacts on type I error, power, and PMD.
In scenarios where multiple historical studies exist within the same population, the framework can be easily expanded to integrate multiple studies into a hybrid study design. The informative prior of $p_c$ can be adjusted to accommodate multiple studies, and the control parameters $w_d$ and $a$ can be tailored for individual studies to address specific circumstances.

{\revone 
The DPP framework involves several adjustable parameters. First, the gated borrowing parameter, $\Delta_{max}$, is usually determined by clinical considerations, such as tumor types. Alternatively, confidence in response rate estimates based on historical data can guide the selection of $\Delta_{max}$. Sensitivity analysis, by varying $\Delta_{max}$, can also be used to assess its impact. Second, the global borrowing parameter $a$ balances tolerable levels of bias and power gain. Sensitivity analysis using different values for $a$, as shown in Table \ref{tab:Sim1Res}, is useful for assessing trade-offs and selecting the optimal parameter value. Third, we used the default values of the $\theta$ and $\eta$ parameters for the GBC and JSD methods based on the literature.\citep{1943_Bhattacharyya,2020_Fujikawa} However, users can select other values based on simulations to optimize performance by balancing bias and power gain.
} 

{\revone The historical control data are incorporated into the modeling by constructing an informative prior. When there is no prior data conflict, i.e., $w_d = 1$, the historical patients are modeled as in-trial, except that we choose to limit the borrowing to a pre-specified amount, $a$. One can also consider the parameter $a$ as a voluntary discount, which results in the overall borrowing $w < 1$, even when there is no prior data conflict.}

The proposed hybrid design relies on a large historical clinical study that includes the control treatment within the same patient population. Given the potential for major protocol deviations in historical studies, it's essential to select patients to ensure their eligibility aligns with the current study criteria. To mitigate bias, historical control patients should be identified and determined prior to analysis, ensuring that the selection process remains independent of study outcomes and reducing the risk of biased patient selection. Moreover, propensity score matching methods can be employed to control for key prognostic factors. Ensuring data quality is paramount in the hybrid design approach. Historical clinical trials conducted by the same sponsor in recent years may offer better consistency in data collection and analytic handling, thereby enhancing the reliability of the historical control data.

{\revone The proposed DPP framework can be extended to continuous and survival endpoints. Traditional power priors are applicable to continuous outcomes\citep{2000_Power_Chen}. Using conjugate prior distributions that facilitate integration with respect to posterior distributions, such as the normal-gamma distribution\citep{1994_Bayesian_Bernardo}, the DPP framework can be adapted for continuous outcomes. Other prior distributions may require numerical integration, Monte Carlo methods, or MCMC techniques. Additionally, several recent publications have studied Bayesian hybrid designs for time-to-event endpoints. For instance, the Bayesian semiparametric meta-analytic predictive prior has been extended to handle time-to-event outcomes\citep{2023_BEATS_Bi}. Hybrid control methods based on Bayesian dynamic borrowing through commensurate priors and frequentist methods using propensity score matching and weighting have also been evaluated for time-to-event outcomes\citep{2024_Evaluating_Wang}. However, survival endpoints require specific considerations, such as handling the baseline hazard function and addressing complications due to a small number of events in early-phase trials. Extending the proposed DPP framework to handle continuous and survival endpoints will be explored in future research.}

While the hybrid design framework theoretically allows for the incorporation of real-world data (RWD), challenges such as inadequate data collection and lack of source data verification may arise. When considering RWD for building the hybrid design, careful examination of feasibility and understanding of associated risks are crucial prior to implementation. Assessing the completeness, accuracy, and reliability of the RWD source is essential to ensure the validity of the analysis results. A recent example of eflornithine's approval in 2023 \citep{2023_FDA_eflornithine} based on a study with external control synthesized from a historical study, following a favorable vote in an oncology drug advisory committee meeting for high-risk neuroblastoma \citep{2023_FDA_ODAC}, has sparked significant interest in the pharmaceutical industry in study designs that can incorporate historical clinical trials. It's important to note that the proposed hybrid design DPP framework is not intended for regulatory filing purposes. Instead, regulatory guidance documents on externally controlled trials (ECTs) should be consulted when considering regulatory filing purposes \citep{2023_FDA_ECA, 2023_EMA_Reflection, 2001_ICH}. The R package and code included in this work are available at \hyperlink{https://github.com/phe9480/BayesianHybridDesign}{https://github.com/phe9480/BayesianHybridDesign}. Legal Note:  Contributions by the authors are solely their own and are not intended to express the views of their organizations.

\section*{Conflict of Interest and Data Availability Statement}
The authors have no conflicts of interest to declare. Data sharing not applicable to this article as no datasets were generated or analyzed during the current study.

%\section*{Reference}

\bibliographystyle{ama.bst}
\bibliography{sample.bib}

%\begin{enumerate}
%  \item Turner NC, Oliveira M, Howell SJ, Dalenc F, Cortes J, Gomez Moreno HL, Hu X, Jhaveri K, Krivorotko P, Loibl S, Morales Murillo S, Okera M, Park YH, Sohn J, Toi M, Tokunaga E, Yousef S, Zhukova L, de Bruin EC, Grinsted L, Schiavon G, Foxley A, Rugo HS; CAPItello-291 Study Group. Capivasertib in Hormone Receptor-Positive Advanced Breast Cancer. N Engl J Med. 2023 Jun 1;388(22):2058-2070. doi: 10.1056/NEJMoa2214131. PMID: 37256976.
%  \item Alivia Crist (2022) New cancer treatment showing 100\% success rate. Morning News.  https://fox59.com/morning-news/new-cancer-treatment-showing-100-success-rate/
%  	\end{enumerate}

\newpage

% Resetting section numbering for supplementary material
\renewcommand{\thesection}{S\arabic{section}}
\renewcommand{\thetable}{S\arabic{table}}
\renewcommand{\thefigure}{S\arabic{figure}}
\renewcommand{\theequation}{S\arabic{equation}}

\setcounter{section}{0} % Reset section counter
\setcounter{table}{0}   % Reset table counter
\setcounter{figure}{0}  % Reset figure counter
\setcounter{equation}{0} % Reset equation counter

\section{Supplementary Materials: Empirical Bayes Method}
       
        \begin{eqnarray}
          w_d &=& \argmax_{w_d\in (0, 1)}\left\{\int_0^1 p(Y_{c}|p_c) p(p_c|Y_{ch}) d p_c\right\}\nonumber\\ 
          &=& \argmax_{w_d\in (0, 1)}\left\{\int_0^1 p(Y_{c}|p_c) p(p_c|Y_{ch}) d p_c\right\}\nonumber\\  % add transition in the chat
          &=& \argmax_{w_d\in (0, 1)}\left\{\frac{\int_0^1 p_c^{Y_c}(1-p_c)^{n_c-Y_c}p_c^{a_{oc} + {w_d}Y_{ch}-1}(1-p_c)^{b_{0c} + {w_d}(n_{ch}-Y_{ch})-1} d p_c}{B(a_{oc} + {w_d}Y_{ch}, b_{0c} + {w_d}(n_{ch}-Y_{ch}))} \right\}\nonumber\\
          %&=& \max_{w_d\in (0, 1)}\left\{\frac{\int_0^1 p_c^{a_{oc} + Y_c+{w_d}Y_{ch}-1}(1-p_c)^{b_{0c} + (n_c-Y_c)+{w_d}(n_{ch}-Y_{ch})-1} d p_c}{B(a_{oc} + {w_d}Y_{ch}, b_{0c} + {w_d}(n_{ch}-Y_{ch}))} \right\}\nonumber\\
          &=& \argmax_{w_d\in (0, 1)}\left\{\frac{B(a_{oc} + Y_c + {w_d}Y_{ch}, b_{0c} + (n_c-Y_c)+{w_d}(n_{ch}-Y_{ch})}{B(a_{oc} + {w_d}Y_{ch}, b_{0c} + {w_d}(n_{ch}-Y_{ch})} \right\}, \nonumber
       \end{eqnarray}
       where $B(a, b)$ is the beta function and calculated as $B(a, b)=\int_0^1 t^{a-1}(1-t)^{b-1}dt$. $w_d$ can be determined by maximizing the ratio of two Beta functions w.r.t $w_d$. 

%    \subsection{Generalized Bhattacharyya Coefficient (BC)}
%    The $w_d$ given $\theta$ parameter that be expressed as the expectations as 
%        \begin{eqnarray}
%    w_d &=& \frac{1}{2}\left[\int_0^1 \left(\frac{f_{ch}}{f_c}\right)^{\theta}f_cdx + \int_0^1 \left(\frac{f_{c}}{f_{ch}}\right)^{\theta}f_{ch}dx\right] \nonumber\\
%        &=& \frac{1}{2}\left[ E_c\left(\frac{f_{ch}}{f_c}\right)^{\theta} + E_{ch}\left(\frac{f_{c}}{f_{ch}}\right)^{\theta}\right].  \nonumber  
%  \end{eqnarray}

\newpage

\section{Supplementary Materials: Additional Information for the Example}
\begin{figure}[ht]
\centering
\includegraphics[height=5in]{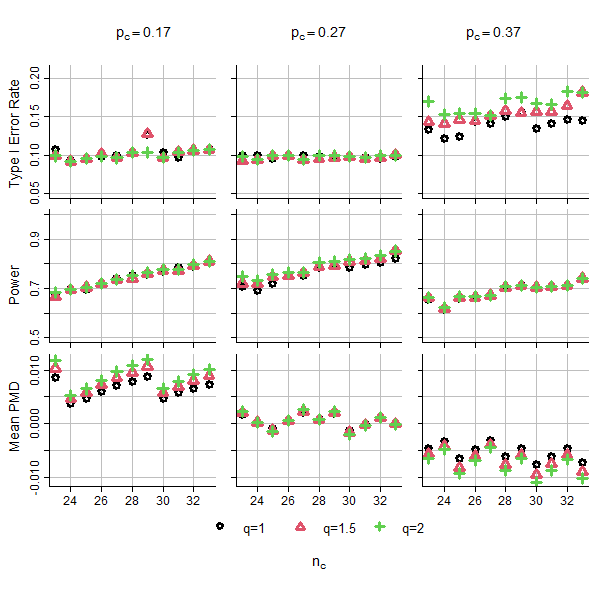}
\caption{Type I error rate, power, and mean PMD using Bayesian $P$ dynamic borrowing method. Calibration is performed assuming the concurrent control and historical control have the same response rate $(p_c=p_{ch}=0.27)$ for one-sided type I error of 0.1. Maximum amount of borrowing $n_{ch,e}=qn_c$.}
\label{fig:exampleBp}
\end{figure}
    
\begin{figure}[ht]
\centering
\includegraphics[height=5in]{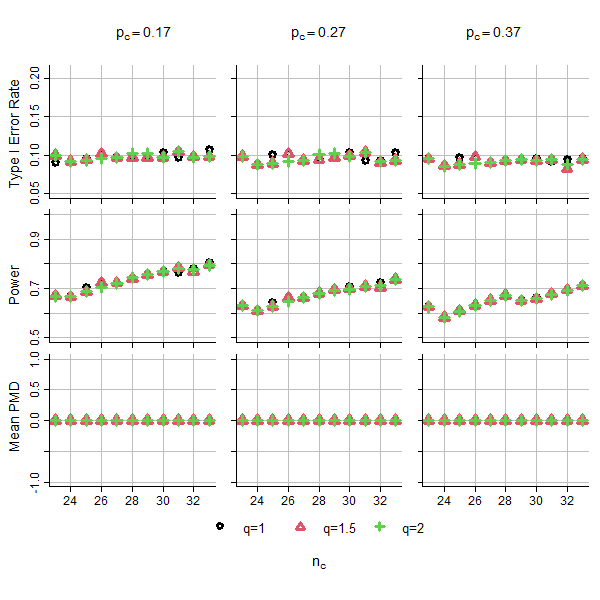}
\caption{Type I error rate, power, and mean PMD without  borrowing from historical data. Calibration is performed assuming the concurrent control and historical control have the same response rate $(p_c=p_{ch}=0.27)$ for one-sided type I error of 0.1.}
\label{fig:exampleEBNoBorrowing}
\end{figure}

%
%\begin{figure}
%    \centering
%    \begin{subfigure}[b]{0.495\textwidth}
%        \centering
%        \includegraphics[width=\textwidth]{wd_EB_hyper.png}
%        \caption{Empirical Bayes}
%        \label{fig:EB}
%    \end{subfigure}
%    \hfill
%    \begin{subfigure}[b]{0.495\textwidth}
%        \centering
%        \includegraphics[width=\textwidth]{wd_BP_hyper.png}
%        \caption{Bayesian $P$}
%        \label{fig:Bp}
%    \end{subfigure}
%    \\
%    \begin{subfigure}[b]{0.495\textwidth}
%        \centering
%        \includegraphics[width=\textwidth]{wd_theta_hyper.png}
%        \caption{Generalized BC}
%        \label{fig:GBC}
%    \end{subfigure}
%    \hfill
%    \begin{subfigure}[b]{0.495\textwidth}
%        \centering
%        \includegraphics[width=\textwidth]{wd_JSD_hyper.png}
%        \caption{Jensen-Shannon Divergence (JSD)}
%        \label{fig:JSD}
%    \end{subfigure}
%    \caption{Sensitivity of dynamic borrowing parameters ($w_d$) to the hyperprior for $p_c$}
%    \label{sfig:wd}
%\end{figure}
%

\begin{table}[ht]
\begin{center}
\caption{Impacts of the Gated Borrowing Parameters ($\Delta_{max}=0.05, 0.1, 0.15$) in the proposed DPP Framework on Type I Error, Power, mean PMD and sd(PMD).}\label{SuppTab:Sim1Res}
{\scriptsize
\begin{tabular}{lr|rrr|rrr|rrr|}
\hline
&& \multicolumn{3}{c}{\makecell{Dynamic power prior \\ $\Delta_{max}=0.05$} }& \multicolumn{3}{c}{\makecell{Dynamic power prior \\ $\Delta_{max}=0.10$}}& \multicolumn{3}{c}{\makecell{Dynamic power prior \\ $\Delta_{max}=0.15$}}\\ \cline{3-11}
  &$p_{c}\backslash n_{ch,e}$& 45 & 90 & 180 & 45 & 90 & 180 & 45 & 90 & 180 \\ \hline

  \hline
 \multirow{7}*{\makecell{Type I\\error\\rate}}&0.15& 0.101 & 0.095 & 0.095 & 0.103 & 0.118 & 0.103 & 0.116 & 0.116 & 0.112 
\\ 
   &0.20& 0.096 & 0.087 & 0.087 & 0.104 & 0.106 & 0.102 & 0.098 & 0.097 & 0.092 
\\ 
   &0.25& 0.095 & 0.089 & 0.089 & 0.096 & 0.090 & 0.087 & 0.090 & 0.084 & 0.076 
\\ 
   &0.30& 0.100 & 0.096 & 0.099 & 0.099 & 0.099 & 0.098 & 0.099 & 0.097 & 0.099 
\\ 
   &0.35& 0.108 & 0.111 & 0.120 & 0.127 & 0.146 & 0.160 & 0.135 & 0.149 & 0.175 
\\ 
   &0.40& 0.107 & 0.113 & 0.122 & 0.150 & 0.187 & 0.205 & 0.172 & 0.201 & 0.245 
\\ 
   &0.45& 0.097 & 0.099 & 0.103 & 0.133 & 0.170 & 0.165 & 0.172 & 0.199 & 0.240 
\\ 
   \hline
 \multirow{7}*{Power}&0.15
& 0.821 & 0.811 & 0.811 & 0.821 & 0.810 & 0.803 & 0.812 & 0.800 & 0.779 
\\ 
   &0.20
& 0.790 & 0.781 & 0.782 & 0.782 & 0.772 & 0.758 & 0.780 & 0.770 & 0.752 
\\ 
   &0.25
& 0.777 & 0.774 & 0.781 & 0.794 & 0.804 & 0.799 & 0.800 & 0.806 & 0.814 
\\ 
   &0.30
& 0.759 & 0.764 & 0.772 & 0.811 & 0.842 & 0.843 & 0.836 & 0.856 & 0.880 
\\ 
   &0.35
& 0.729 & 0.731 & 0.735 & 0.779 & 0.823 & 0.799 & 0.833 & 0.855 & 0.879 
\\ 
   &0.40
& 0.712 & 0.712 & 0.713 & 0.733 & 0.791 & 0.738 & 0.804 & 0.821 & 0.830 
\\ 
   &0.45& 0.710 & 0.709 & 0.709 & 0.720 & 0.779 & 0.720 & 0.785 & 0.791 & 0.787 
\\ 
   \hline
   \multirow{7}*{\makecell{Mean \\ PMD}}&0.15
& 0.000 & 0.000 & 0.000 & 0.004 & 0.006 & 0.009 & 0.007 & 0.011 & 0.017 
\\ 
   &0.20
& 0.001 & 0.002 & 0.002 & 0.009 & 0.014 & 0.019 & 0.011 & 0.018 & 0.026 
\\ 
   &0.25
& 0.001 & 0.002 & 0.002 & 0.008 & 0.012 & 0.016 & 0.009 & 0.014 & 0.019 
\\ 
   &0.30
& 0.000 & 0.000 & 0.000 & 0.000 & 0.001 & 0.002 & -0.001 & -0.001 & -0.001 
\\ 
   &0.35
& -0.001 & -0.001 & -0.002 & -0.007 & -0.009 & -0.011 & -0.011 & -0.015 & -0.021 
\\ 
   &0.40
& -0.001 & -0.001 & -0.002 & -0.008 & -0.011 & -0.013 & -0.014 & -0.020 & -0.028 
\\ 
   &0.45& -0.001 & -0.001 & -0.001 & -0.005 & -0.006 & -0.008 & -0.010 & -0.016 & -0.023 
\\ 
   \hline
 \multirow{7}*{\makecell{sd(PMD)}}&0.15
& 0.002 & 0.003 & 0.004 & 0.009 & 0.013 & 0.018 & 0.008 & 0.013 & 0.018 
\\ 
   &0.20
& 0.005 & 0.007 & 0.008 & 0.012 & 0.017 & 0.023 & 0.011 & 0.016 & 0.020 
\\ 
   &0.25
& 0.007 & 0.010 & 0.012 & 0.016 & 0.023 & 0.029 & 0.016 & 0.023 & 0.029 
\\ 
   &0.30
& 0.008 & 0.011 & 0.014 & 0.019 & 0.026 & 0.033 & 0.020 & 0.028 & 0.036 
\\ 
   &0.35
& 0.008 & 0.010 & 0.012 & 0.018 & 0.024 & 0.030 & 0.018 & 0.025 & 0.033 
\\ 
   &0.40
& 0.006 & 0.007 & 0.009 & 0.015 & 0.021 & 0.025 & 0.015 & 0.021 & 0.027 
\\ 
   &0.45& 0.003 & 0.005 & 0.006 & 0.012 & 0.016 & 0.019 & 0.013 & 0.018 & 0.025 \\ 
   \hline
\end{tabular}
} 
\end{center}
{Notes: For DPP methods, $w_d$ is determined by empirical Bayes approach.}
%{Notes: $p_{ch}$ is the historical control ORR. $w_d$ in dynamic power prior was determined by Bayesian p-value. The several choices of global parameters $a=n_{ch,e}/180$ are used to represent different amount of information borrowed from external control.}
\end{table}

\end{document}